\documentclass{article}

\usepackage{arxiv}

\usepackage[utf8]{inputenc} 
\usepackage[T1]{fontenc}    
\usepackage{hyperref}       
\usepackage{url}            
\usepackage{booktabs}       
\usepackage{amsfonts}       
\usepackage{nicefrac}       
\usepackage{microtype}      
\usepackage{lipsum}
\usepackage{graphicx}
\graphicspath{ {./images/} }

\title{A Scalable and Automated Framework for Tracking the likely Adoption of Emerging Technologies}

\author{
 Lowri Williams\\
  School of Computer Science \& and Informatics\\
  Cardiff University \\
  UK \\
  \texttt{WilliamsL10@cardiff.ac.uk} \\
   \And
 Eirini Anthi\\
  School of Computer Science \& and Informatics\\
  Cardiff University \\
  UK \\
  \texttt{AnthiES@cardiff.ac.uk} \\
  \And
 Pete Burnap \\
  School of Computer Science \& and Informatics\\
  Cardiff University \\
  UK \\
  \texttt{BurnapP@cardiff.ac.uk} \\
}

\begin{document}
\maketitle
\begin{abstract}
While new technologies are expected to revolutionise and become game-changers in improving the efficiencies and practises of our daily lives, it is also critical to investigate and understand the barriers and opportunities faced by their adopters. Such findings can serve as an additional feature in the decision-making process when analysing the risks, costs, and benefits of adopting an emerging technology in a particular setting. Although several studies have attempted to perform such investigations, these approaches adopt a qualitative data collection methodology which is limited in terms of the size of the targeted participant group and is associated with a significant manual overhead when transcribing and inferring results. This paper presents a scalable and automated framework for tracking likely adoption and/or rejection of new technologies from a large landscape of adopters. In particular, a large corpus of social media texts containing references to emerging technologies was compiled. Text mining techniques were applied to extract sentiments expressed towards technology aspects. In the context of the problem definition herein, we hypothesise that the expression of positive sentiment infers an increase in the likelihood of impacting a technology user's acceptance to adopt, integrate, and/or use the technology, and negative sentiment infers an increase in the likelihood of impacting the rejection of emerging technologies by adopters. To quantitatively test our hypothesis, a ground truth analysis was performed to validate that the sentiment captured by the text mining approach is comparable to the results given by human annotators when asked to label whether such texts positively or negatively impact their outlook towards adopting an emerging technology. The collected annotations demonstrated comparable results to those of the text mining approach, illustrating that automatically extracted sentiment expressed towards technologies are useful features in understanding the landscape faced by technology adopters, as well as serving as an important decision-making component when, for example, recognising shifts in user behaviours, new demands, and emerging uncertainties.
\end{abstract}


\section{Introduction}
Technological change is revolutionising the way we lead our daily lives. From the way we work to our own homes, the integration of new technologies, such as \textit{5G, Internet of Things (IoT) devices}, and \textit{Artificial Intelligence (AI)}, enhances our productivity and efficiencies \cite{jackson2022}. However, while new technologies evolve and are expected to revolutionise practices, the industry is facing various barriers in terms of their adoption and implementation. The technology adoption process is affected by aspects such as the availability and quality of hardware/software, organisational role models, available financial resources and funding, organisational support, staff development, attitudes, technical support, and time to learn new technology \cite{rogers2000barriers}.

Understanding the barriers and advantages faced by technology adopters can be a key feature which impacts essential business decision-making components, such as recognising shifts in user behaviours, new demands, and emerging uncertainties. As such, this information can significantly aid in the development of an adequate response, such as developing a technology for addressing the new requirements in an efficient way \cite{kucharavy2008technological} or developing new organisational strategies. 

Several studies have focused on investigating the current positive experiences and barriers of adopting emerging technologies in various settings, such as in educational institutions (e.g. \cite{bacow2012barriers}), healthcare (e.g. \cite{christodoulakis2017barriers}), smart-medium enterprises (e.g. \cite{kapurubandara2006barriers}), and by older adults (e.g. \cite{yusif2016older}). Such studies often rely on qualitative data collection methods, such as focus groups and interviews. However, this approach is associated with several limitations, including collecting data from a small and targeted sample size which limits the variance and bias in responses, the significant overhead associated with recruiting participants, organising interviews, and manually transcribing and inferring results. To gain an understanding of the landscape surrounding emerging technologies from a larger sample of adopters and across different settings, a wider analysis is needed. 

To extract such data at scale, automated techniques are needed to collect and programmatically extract relevant information from publicly available sources. Such sources may include online social media platforms (e.g. Twitter) in which users can publish their content, presenting a wealth of information surrounding their opinions and experiences \cite{burnap2016us}. To automatically extract and process large volumes of texts originating from diverse sources, text mining techniques may be used \cite{williams2015}. In particular, sentiment analysis, often referred to as opinion mining, aims to automatically extract and classify sentiments and/or emotions expressed in text \cite{liu2010sentiment, munezero2014}. In the context of the aforementioned problem definition, we hypothesise that the expression of positive sentiment infers opportunities and positive experiences surrounding emerging technologies, which, therefore, increases the likelihood of impacting a technology user's acceptance to adopt, integrate, and/or use the technology. Conversely, we hypothesise that the expression of negative sentiment infers the obstacles and barriers caused and faced by such technologies, which, therefore, increases the likelihood of impacting the rejection of emerging technologies by adopters.

To the best of our knowledge, this paper presents the first scalable and automated framework towards tracking the likely adoption of emerging technologies. Such framework is powered by the automatic collection and analysis of social media discourse containing references to emerging technologies from a large landscape of adopters. The main contributions of the work presented herein are as follows:

\begin{itemize}
\item The extraction of aspects relating to a range of emerging technologies from social media discourse over a period of time.

\item The classification of sentiment expressed towards such technologies, indicating the positive and negative outlook of users towards adopting them.

\item A ground truth analysis to validate the hypothesis that the sentiment captured by the text mining approach is comparable to the results given by human annotators when asked to label whether such texts positively or negatively impact their outlook towards adopting an emerging technology. 

\item A scalable and automated framework for tracking the likely adoption and/or rejection of new technologies. This information serves as an important decision-making component when, for example, recognising shifts in user behaviours, new demands, and emerging uncertainties.

\item Resources that can further support research into narratives surrounding emerging technologies, such as a large corpus of social media discourse covering five year's worth of data.  
\end{itemize} 

The study was designed as follows: 1) compile a corpus of texts containing references to emerging technologies over a time period, 2) pre-process text responses using traditional Natural Language Processing (NLP) techniques, 3) divide texts based on their publication date, 4) for each dataset in 3), automatically extract technology aspects from text segments, 5) for each dataset in 3), apply a sentiment analysis approach to automatically extract the sentiment expressed towards the identified aspects, and 6) visualise and analyse the results. 

The remainder of this paper is structured as follows: Section \ref{sec:background} presents the related work, Section \ref{sec:datasets} discusses the collection of texts used to support the experiments herein and the techniques used to prepare the data for such experiments, Section \ref{sec:aspect sentiment analysis} discusses aspect-based sentiment analysis and how it was applied to the datasets, Section \ref{sec:results} presents and discusses the results, Section \ref{sec:evaluation} quantitatively evaluates our hypothesis by comparing the automated text mining method against the impact such texts have on technology adopters, Section \ref{sec:conclusion} concludes the paper, and finally, Section \ref{sec:future work} discusses future work.

\section{Related Work}\label{sec:background}

Several studies have explored the factors influencing technology adoption in several different settings such as healthcare, education, smart-medium enterprises, and by older adults. Such studies have often adopted theoretical frameworks, such as Technology Acceptance Model (TAM) \cite{davis1989perceived} and the Unified Theory of Acceptance and Use of Technology (UTAUT) \cite{venkatesh2003user}, to understand and predict the acceptance and adoption of new technologies based on factors such as their perceived usefulness and ease of use. For example, by using such frameworks, Alalwan et al. \cite{alalwan2017factors} and \cite{oliveira2016mobile} et al. explore factors that influence mobile banking uptake and customer uptake and adoption of mobile payment technologies respectively. Likewise, Bhattacherjee \& Park \cite{bhattacherjee2014end} investigate factors that influence end-user migration to cloud computing services. To collect customer and user data, the aforementioned studies use a survey-based methodology.

To derive the barriers faced within a healthcare setting, Sun \& Medaglia \cite{sun2019mapping} investigate the perceived challenges of AI adoption in the public healthcare sector in China. Their study relies on data collected from semi-structured interviews asking a sample group of seven key stakeholder groups open-ended questions focusing on the challenges of AI adoption in healthcare. Similarly, Al-Hadban et al. \cite{al2017barriers} explore the opinions of healthcare professionals using semi-structured interviews to highlight the important factors and issues that influence the adoption of new technologies in the public healthcare sector in Iraq. Their study relies on data collected from a sample group of eight interviewees. They describe their data collection approach which includes producing transcriptions of the audio recordings, interpreting and understanding the general sense of the text to form themes, and validating the accuracy of their findings as a time-consuming process. Poon et al. \cite{poon2006assessing} assess the level of healthcare information technology adoption in the United States by also implementing semi-structured interviews with 52 participants from eight key stakeholder groups. They describe that one of the key limitations of their study is the responder biases caused by selecting participants based on their access to contacts. 

To derive barriers and positive experiences of adopting emerging technologies in an educational setting, Jin et al. \cite{jin2022will} also implement their data collection methods by conducting semi-structured interviews with nine instructors and nine students to understand their perceptions of educational \textit{Virtual Reality (VR)} technologies. Dequanter et al. \cite{dequanter2022determinants} examine the factors underlying technology use in older adults with mild cognitive impairments. In their study, over the course of two years, they conducted semi-structured interviews with 16 adults aged 60 and over from a single area in Belgium. Similar to Poon et al. \cite{poon2006assessing}, both Jin et al. \cite{jin2022will} and Dequanter et al. \cite{dequanter2022determinants} describe that one of the key limitations of their study is the responder biases as they believe they recruited participants that were more interested in using \textit{VR} or novel technologies and from one specific geographical area which limits the transferability of the results.

The aforementioned works focus on applying qualitative approaches towards understanding factors which surround emerging technology adopters in several different settings. However, it is evident that such approaches are faced with significant limitations such as recruiting few participants to partake in their studies and the bias in their responses. As a response to such limitations, recent studies have turned to text mining approaches to automatically analyse, and ultimately help understand the factors influencing consumer adoption or rejection of emerging technologies from a large landscape of adopters. For example, Kwarteng et al. \cite{kwarteng2020consumer} investigate applying sentiment analysis to Twitter data to provide insights into consumer perceptions, emotions, and attitudes towards autonomous vehicles. Efuwape et al. \cite{efuwape2022text} investigate the acceptance and adoption of digital collaborative tools for academic planning using sentiment analysis of responses gathered in a poll. Hizam et al. \cite{hizam2022web} employ sentiment analysis to examine the correlations between numerous factors of technology adoption behaviour, such as perceived ease of use, perceived utility, and social impact. The research aims to understand the underlying variables driving Web 3.0 adoption and offers insight on how these factors influence users' decisions to accept or reject these emerging technologies by analysing user-generated content on social media sites. Mardjo \& Choksuchat \cite{mardjo2022hyvadrf} and Caviggioli et al. \cite{caviggioli2020technology} investigate using sentiment analysis to examine the public's perception of adopting Bitcoin. The goal of such studies is to forecast the sentiment of Bitcoin-related tweets, which could influence the cryptocurrency's market behaviour, as well as providing insights into how the public reacts to the adoption of Bitcoin, and how it affects the perception of the adopting companies. Ikram et al. \cite{ikram2016open} investigate how potential adopters perceive specific features of open-source software by examining the sentiment expressed on Twitter.

While the aforementioned studies involve investigating sentiment analysis as an approach towards understanding the factors influencing consumer adoption or rejection of emerging technologies, such studies have primarily focused on specific technologies. Additionally, they lack a ground truth analysis to corroborate the sentiment gathered by the text mining approach, raising worries about the robustness of their findings. As a result, there is an opportunity to expand on existing research by creating a framework that allows for the examination of a greater range of technologies, resulting in a more comprehensive knowledge of the factors influencing their adoption. Furthermore, this framework can be customised to focus on specific sectors or technologies, supporting decision-making processes by identifying shifts in user behaviour, new expectations, and growing uncertainties. This study will not only add to the existing body of knowledge by broadening the scope of analysis and allowing for greater customisation, but it will also provide valuable insights that can inform strategic decisions across industries as they navigate the challenges and opportunities associated with technology adoption.

\section{Data Collection and Preparation}\label{sec:datasets}

To explore online narratives surrounding emerging technologies, textual data was collected from Twitter, a social networking service that enables users to send and read tweets – text messages consisting of up to 280 characters. Snscrape\footnote{https://github.com/JustAnotherArchivist/snscrape}, a Python scraper for social networking services, was used to scrape English tweets. To facilitate the concept of the framework presented herein, five year's worth of tweets published between 01/01/2016 and 31/12/2021 was collected as it aligns with the increase in the adoption of one of the most popular emerging technologies, the Internet of Things (IoT) \cite{iot}.

The IoT refers to the collection of smart devices which have ubiquitous connectivity, allowing them to communicate and exchange information with other technologies \cite{anthi2019supervised}. As more devices connect to the internet and to one another, the IoT is an emerging technology which is considered as being amongst the biggest disruptors, particularly for companies across industries, due to their ability to innovate and develop new products and services, increase productivity with higher levels of performance, improve inventory management, and allow greater access to consumer data to observe patterns and behaviours for continued product and service enhancements \cite{pwc}. In this case, in this paper, tweets published between the aforementioned dates were collected based on the presences of the hashtags ``IoT'' or ``Internet of Things''. A total of 4,520,934 tweets containing the aforementioned keywords were collected and divided into datasets based on the month and year they were published. The dataset with the most tweets (92,290) was reported in November 2017, with December 2021 reporting the fewest tweets (29,793). No retweets or quote retweets were collected; only self-authored tweets were to avoid duplicated data. 

The dataset is available on GitHub\footnote{INSERT IF ACCEPTED} and is released in compliance with Twitter's Terms and Conditions, under which we are unable to publicly release the text of the collected tweets. We are, therefore, releasing the tweet IDs, which are unique identifiers tied to specific tweets. The tweet IDs can be used by researchers to query Twitter's API and obtain the complete tweet object, including tweet content (text, URLs, hashtags, etc.) and authors' metadata. 

The data preparation and analysis in this study was conducted using Python (version 3.7.2). For text pre-processing, the following standard NLP techniques were applied:

\begin{itemize}
    \item Converting text to lowercase.
    
    \item Removing mentioned usernames, hashtags, and URLs using Python's regular expression package, RegEx (version 2020.9.27).
    
    \item To remove bias from the analysis, the keywords (i.e. ``IoT'' and ``Internet of Things'')  used to scrape tweets were also removed.
\end{itemize}

\section{Aspect-Based Sentiment Analysis}\label{sec:aspect sentiment analysis}

Aspect-based sentiment analysis is a text mining technique which aims to identify aspects (e.g. foods, sports, countries) and the sentiment (the subjective part of an opinion) and/or emotion (the projections or display of a feeling) expressed towards them. This technique is often achieved by performing: 

\begin{itemize}
    
    \item \textbf{Aspect extraction} - aims to automatically identify and extract specific entities and/or properties of entities in text \cite{thet2010aspect}.
    
    \item \textbf{Sentiment analysis} - often referred to as opinion mining, sentiment analysis aims to automatically extract and classify sentiments and/or emotions expressed in text \cite{liu2010sentiment, munezero2014}.
    
\end{itemize}

The following Sections further present how aspects relating to emerging technologies and the sentiment expressed towards them were extracted from text in more detail, as well as the results following the application of such techniques on the dataset presented in Section \ref{sec:datasets}.

\subsection{Aspect Extraction} \label{sec:aspect_extraction}

There are various methods by which aspects can be extracted from text. For example, aspect extraction may be achieved using topic modelling, a text mining technique used to identify and extract salient concepts or themes referred to as ``topics'' distributed across a collection of texts \cite{kang2019analysis}. The output from applying topic modelling is commonly a set of the top most co-occurring terms appearing in each topic \cite{jacobi2016quantitative, greene2014many}. However, some of the issues with applying topic modelling methods (e.g. spaCy \cite{spacy}, Gensim \cite{Gensim}) to achieve aspect extraction are that the pre-trained models provided by these libraries are not specific to emerging technologies and may not be able to recognise or accurately identify new or specialised terms related to this field. In addition, there is often manual overhead associated with interpreting aspects extracted by such methods. For example, ``\textit{car, power, light, drive, engine, turn}'' may infer topics surrounding \textit{Vehicles}, and ``\textit{game, team, play, win, run, score}'' may infer \textit{Sports}. Another similar method for extracting aspects is named entity recognition, a technique for extracting named entities, such as names, geographic locations, ages, addresses, phone numbers, etc. from the text. However, both topic modelling and named entity recognition methods may over-generalise the aspects extracted from texts, in turn, losing finer-grained entities. In addition, challenges may occur when topic modelling outputs present irrelevant terms, such as ``\textit{car, power, light, cake, baking, chocolate}'', where the overall aspect cannot be defined.

In this case, for each pre-processed dataset described in Section \ref{sec:datasets}, a simple direct string matching approach was applied to automatically extract aspects that could be mapped against the mapping reference of the Cybersecurity Body of Knowledge (CyBOK) \cite{cybok}, a resource which provides an index of cybersecurity referenced terms, including emerging technology terms. Of the 13,037 terms available in CyBOK v1.3.0, 3,911 were extracted from the corpus, with one tweet containing a maximum of 20 terms, 514,458 tweets containing a minimum of 1 term, and 3,472,358 tweets containing no terms. Table \ref{tab:preprocessing_text} reports examples of the CyBOK aspects extracted from tweets.

\begin{table}
\centering
\caption{Examples of tweets mapped to CyBOK terms}
\label{tab:preprocessing_text}
\begin{tabular}{|l|l|}
\hline
\multicolumn{1}{|c|}{Tweet}                                                                                                                                                                                                                                                                        & \multicolumn{1}{c|}{Extracted Aspect}  \\ \hline
\begin{tabular}[c]{@{}l@{}}One of our engineers was at the Google \\ Cloud On-Board roadshow this morning. \\ It's great to hear we're being described as \\ an industry leader in cloud native platform \\ delivery!\end{tabular}                                                                 & 'google', 'cloud', 'cloud native'      \\ \hline
\begin{tabular}[c]{@{}l@{}}5 of the best Alexa-enabled devices for \\ automation.\end{tabular}                                                                                                                                                                                                     & 'devices', 'automation'                \\ \hline
\begin{tabular}[c]{@{}l@{}}How the growth of IoT is changing data \\ management.\end{tabular}                                                                                                                                                                                                      & 'data management'                      \\ \hline
\begin{tabular}[c]{@{}l@{}}Challenge in talent part of implementing: \\ marrying domain knowledge of software \\ and hardware engineers; finding people \\ who can wear many hats; competitiveness\\  of the data science space; realize what's \\ possible is changing all the time.\end{tabular} & 'software', 'hardware', 'data science' \\ \hline
\begin{tabular}[c]{@{}l@{}}Mobile re-emerges as revolutionizing tech \\ behind virtual reality, machine learning.\end{tabular}                                                                                                                                                                     & 'mobile', 'machine learning'           \\ \hline
\end{tabular}
\end{table}

\begin{figure}[ht]
  \centering
  \vspace{3mm} 
  \includegraphics[width=0.5\textwidth]{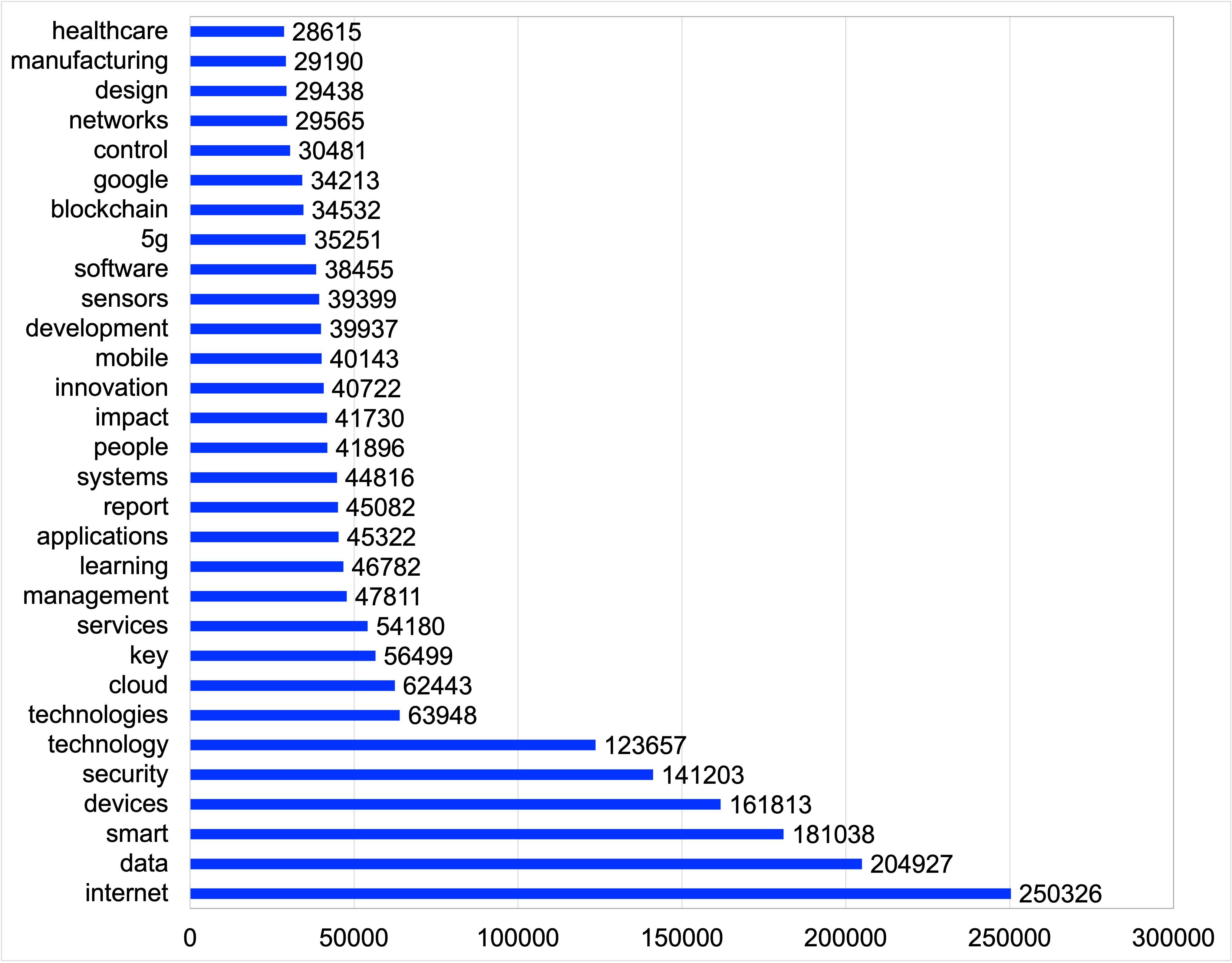}
\caption{Top 30 most frequently referenced CyBOK terms used across the dataset}
\label{technology_distribution}
\end{figure}

Having removed the keywords used to scrape the tweets, Figure \ref{technology_distribution} reports the distribution of extracted terms from CyBOK across the dataset.

\subsection{Sentiment Analysis}\label{sec:sentimentanalysis}

Sentiment analysis, often referred to as opinion mining, aims to automatically extract and classify sentiments and/or emotions expressed in text \cite{liu2010, munezero2014}. Most research activities focus on sentiment classification, which classifies a text segment (e.g. phrase, sentence or paragraph) in terms of its polarity: positive, negative or neutral. Various techniques and methodologies have been developed to address the automatic identification and extraction of sentiment expressed within free text. The two main approaches are the rule-based approach, which relies on predefined lexicons of opinionated words and calculates a sentiment score based on the number of positive or negative words in the text, and automated sentiment analysis, which relies on training a machine learning algorithm to classify sentiment based on both the words in the text and their order.

While there exists a variety of sentiment analysis methods, in the work herein, Valence Aware Dictionary and Sentiment Reasoner (VADER) \cite{hutto2014vader}, a lexicon-based sentiment analysis tool was employed. VADER not only aligns with other relevant studies in the field (e.g. \cite{kwarteng2020consumer, efuwape2022text, hizam2022web, mardjo2022hyvadrf}) and therefore ensures consistency and comparability with existing research, but it is also specifically tuned to classify sentiment expressed in social media language, such as the dataset collated in Section \ref{sec:datasets}. VADER takes into account various features of social media language, such as the use of exclamation marks, capitalisation, degree modifiers, conjunctions, emojis, slang words, and acronyms, which can all impact the sentiment intensity and polarity of a tweet. For example, the use of an exclamation mark increases the magnitude of the sentiment intensity without modifying the semantic orientation, while capitalising a sentiment-relevant word in the presence of non-capitalised words increases the magnitude of the sentiment intensity. Given the complexity of social media language and the various features that can impact sentiment analysis, using VADER to extract the sentiment expressed in the dataset collated herein allows for more accurate and nuanced sentiment analysis of the tweets.

VADER provides a percentage score, which represents the proportion of the text which falls in the positive, negative, or neutral categories. To represent a single uni-dimensional measure of sentiment, VADER also provides a compound score which is computed by summing the valence scores of each word in the lexicon and then normalising the scores to be between -1 (most extreme negative) and 1 (most extreme positive). In the work herein, text segments with a compound score $<=$ -0.05 were considered as expressing a negative sentiment, those with a score $>$ -0.05 and $<$ 0.05 were considered as expressing neutral sentiment, and those with a score $>=$ 0.05 were considered to express positive sentiment. For example, the tweet \textit{`No fear that a hacker can get access to your camera or thermostat or other electronic devices. Your privacy is 100\% protected because the technology is inside your electronics and not located on any server across the world.'} achieved a compound score of 0.6734 and was therefore assigned a positive sentiment. For each extracted aspect, the final sentiment class was assigned by taking the polarity with the highest average compound score. For example, for the aspect \textit{4G network}, in December 2018, the average compound scores were as follows: positive = 0.1027, negative = 0.58, and neutral = 0. In this case, the overall sentiment assigned for \textit{4G network} during this time was negative.

\section{Results and Discussion}\label{sec:results}

\begin{figure}
  \centering
  \vspace{3mm} 
  \includegraphics[width=1\textwidth]{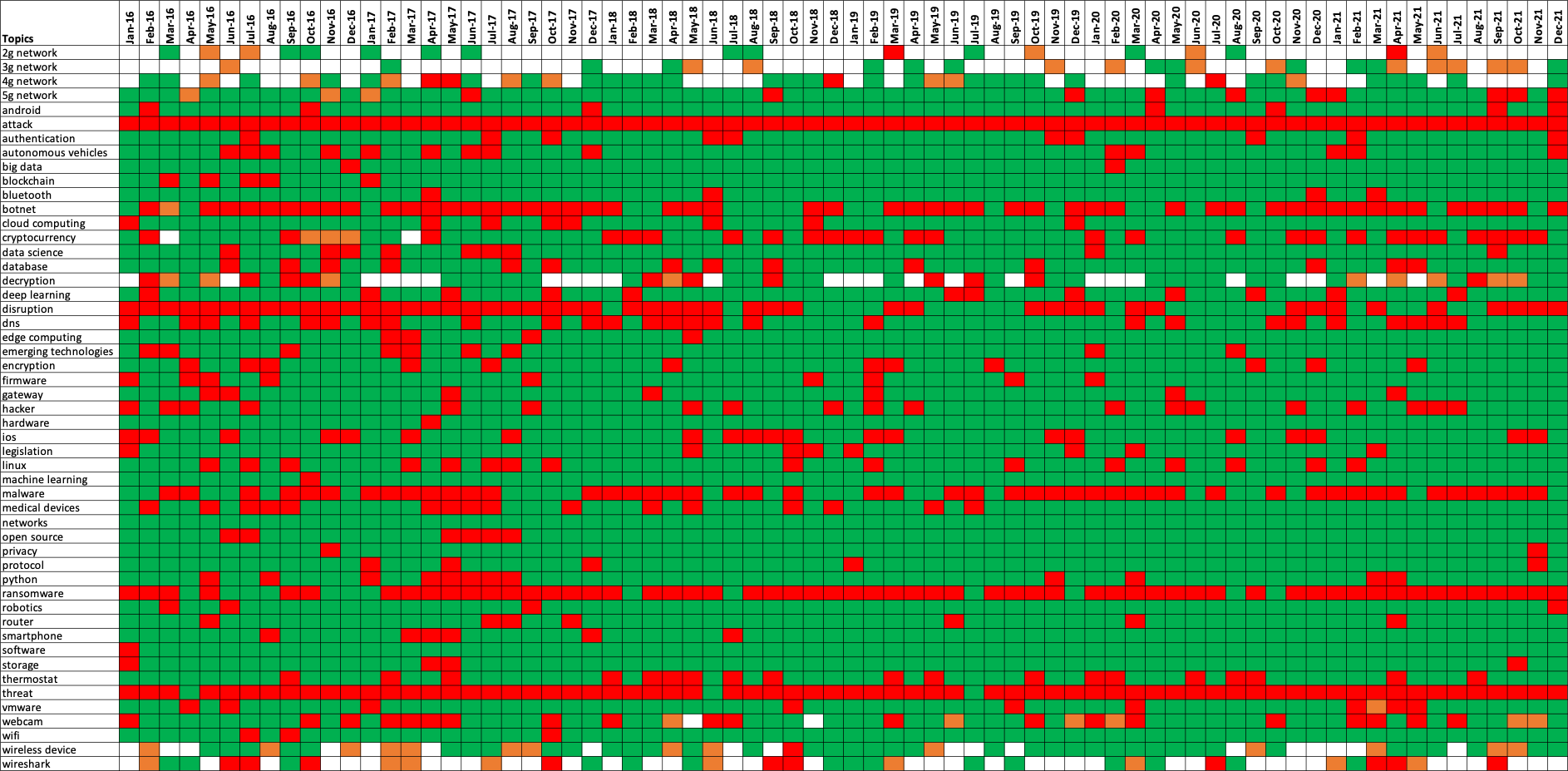}
\caption{Chronological aspect-based analysis results across an excerpt of emerging technologies from the whole dataset (positive sentiment = green, neutral sentiment = orange, negative sentiment = red)}
\label{results}
\end{figure}

Figure \ref{results} reports chronological aspect-based analysis results across an excerpt of emerging technologies across the whole dataset. The figure depicts chronological monthly data, presenting changes in sentiment over the timeline. However, it is possible to refine the data to show daily results to gain more granular information. Despite this, the monthly data is still useful for monitoring outputs from a broader perspective. While the figure may not capture every detail, it provides an overview of trends and changes over time that can be used to inform decision-making and identify potential areas for improvement.

By observing sentiment expressed towards emerging technologies alongside the stages of technology adoption presented by Rogers \cite{rogers2003diffusion} and the various adopter categories (i.e. innovators, early adopters, early majority, late majority, and laggards) presented by Moore \cite{moore1991crossing}, it is possible to gain a better understanding of how shifts in sentiment may influence the adoption or rejection of emerging technologies. During the early stages of \textit{5G network} adoption in April 2020, for instance, negative sentiment was expressed in the form of conspiracy theories and misinformation (e.g., \textit{'Rumours of 5G as the true cause behind COVID-19, communication towers burned...'}). Such data may have a greater effect on laggards, who are typically more risk-averse and resistant to new technologies \cite{rogers2003diffusion}. In contrast, early consumers, who are typically more receptive to innovation and risk \cite{moore1991crossing}, may be more interested in the potential advantages and opportunities of the \textit{5G network}. For example, in November 2021, several tweets (e.g. \textit{'5G network compatibility will make IoT devices better suited for the future as the industry continues to see how the speed 5G provides can make IoT devices preform better'} and \textit{'Microsoft and AT\&T are accelerating the enterprise customer's journey to the edge with 5G'}) expressed an overall positive sentiment. As the \textit{5G network} matures and progresses through the adoption stages, users in the early and late majority stages may become more familiar with the benefits of the technology and their attitudes may change. This change may be indicative of a broader acceptance of the technology, resulting in its increased adoption.

Similarly, when analysing cybersecurity issues such as cyber attacks, various adopter categories may be influenced differently by the sentiment conveyed. In the case of \textit{malware}, the negative sentiment expressed in July 2017 regarding the WORM-RETADUP attack (e.g. \textit{'Information-stealing malware discovered targeting Israeli hospitals'}), innovators and early adopters may view this as a learning opportunity and work to develop more robust security measures. In contrast, the late majority and laggards may be discouraged from adopting these technologies due to cybersecurity concerns. Likewise, the negative sentiment expressed in June 2016 related to the Distributed Denial-of-Service (DDoS) attack involving compromised CCTV cameras could impact adoption patterns across adopter categories in a similar manner (e.g. \textit{`25,000 CCTV cameras hacked to launch \#DDoS attack'} and \textit{'What a way to cause a distraction - 25,000 CCTV cameras hacked to launch DDoS attack'}).

\begin{figure}
  \centering
  \vspace{3mm} 
  \includegraphics[width=1\textwidth]{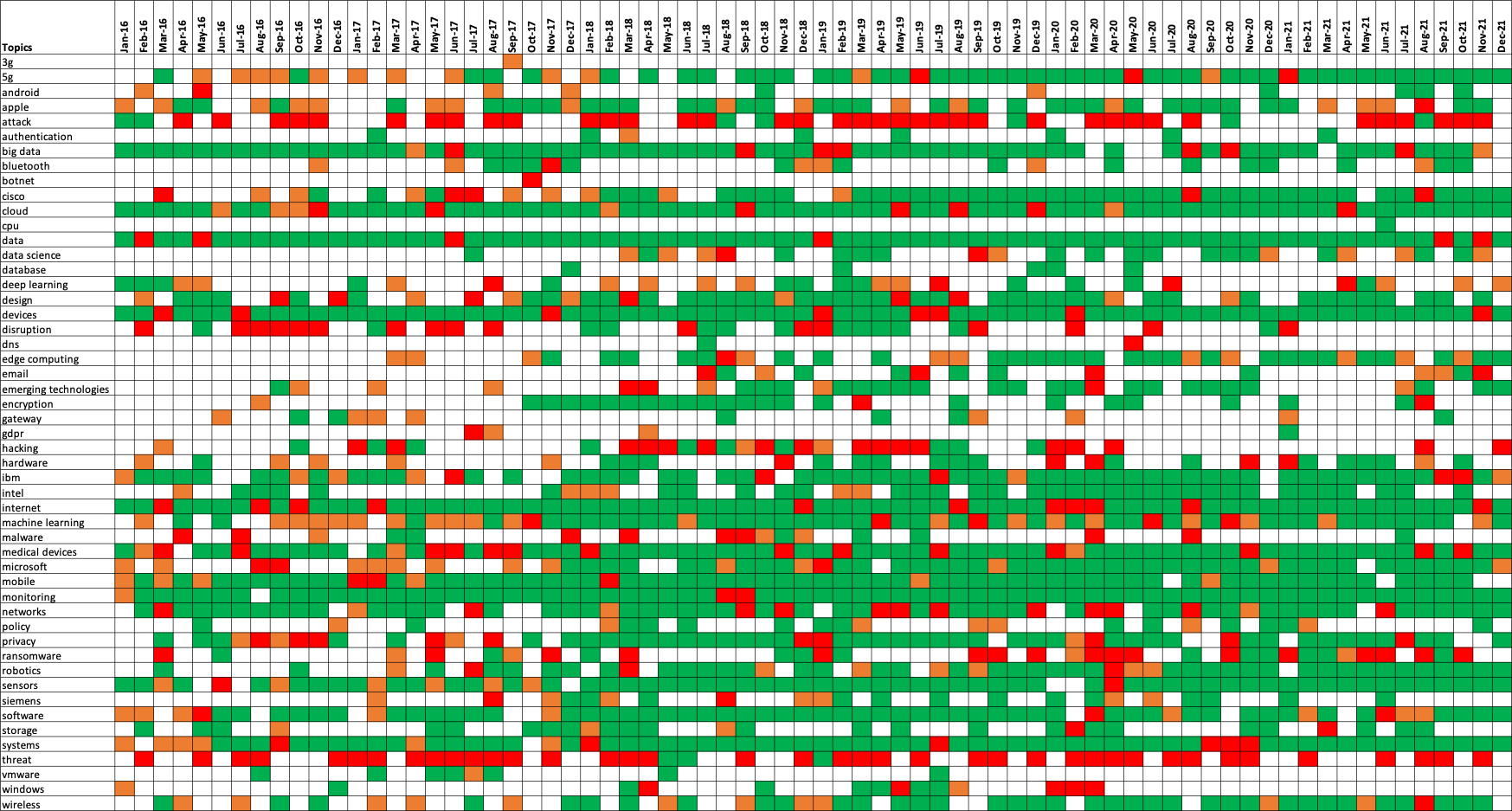}
\caption{Chronological aspect-based analysis results across an excerpt of emerging technologies from discourse relating to healthcare (positive sentiment = green, neutral sentiment = orange, negative sentiment = red)}
\label{healthcare}
\end{figure}

\begin{figure}
  \centering
  \vspace{3mm} 
  \includegraphics[width=1\textwidth]{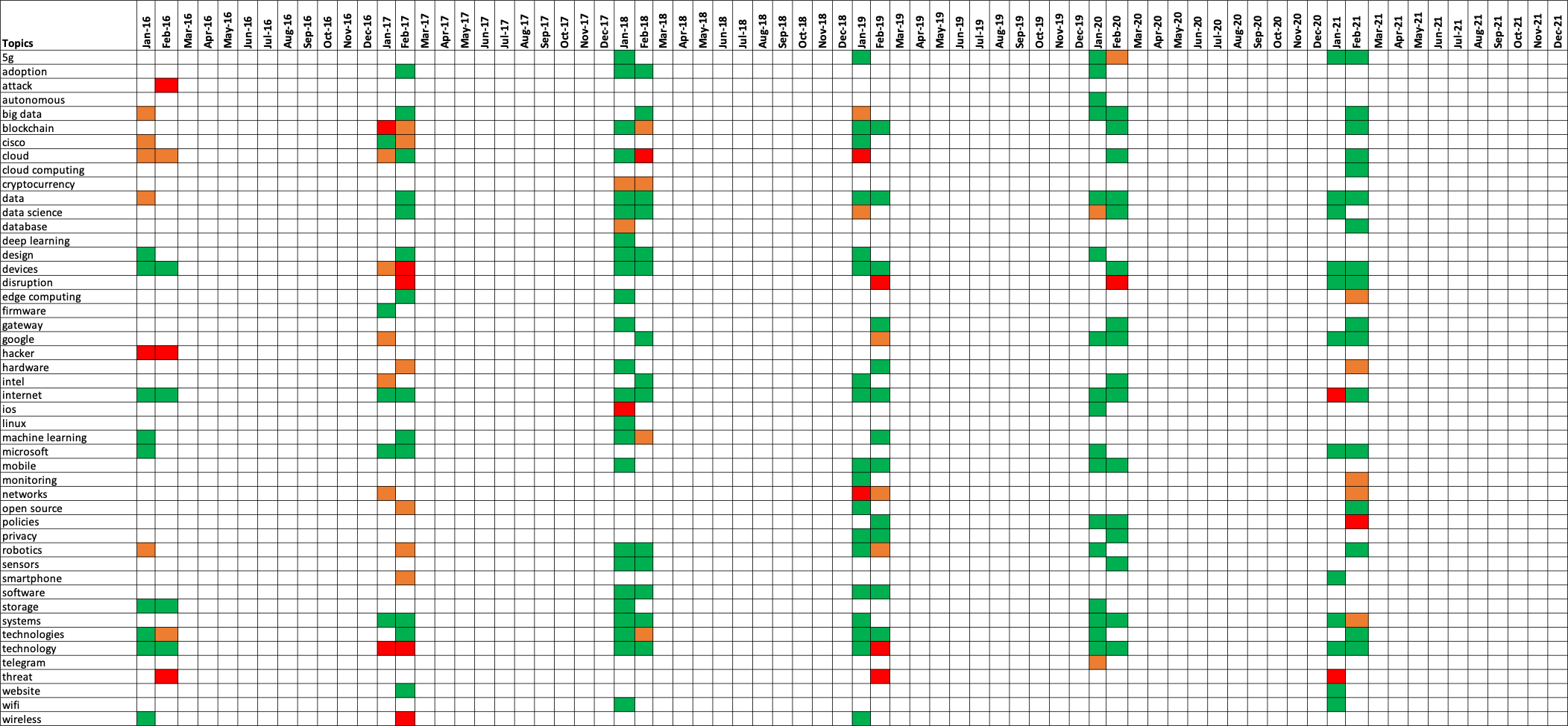}
\caption{Chronological aspect-based analysis results across an excerpt of emerging technologies from discourse relating to education (positive sentiment = green, neutral sentiment = orange, negative sentiment = red)}
\label{education}
\end{figure}

By refining the data, the framework can be customised to concentrate on specific industries, which enhances its flexibility and makes it a dynamic solution that can adapt to the unique needs of different sectors. For example, Figures \ref{healthcare} and \ref{education} report chronological aspect-based analysis results across an excerpt of emerging technologies from discourse relating to healthcare and education respectively. 

In the health sector, in August 2017an overall negative sentiment was expressed towards Siemens' medical molecular imaging systems (e.g. \textit{'This type of vulnerability in healthcare is not unique to Siemens'}) as an alert warning was issued when publicly available exploits were identified that could allow an attacker to remotely execute damaging code or compromise the safety of their systems \cite{siemens}. Innovators and early adopters might view the security vulnerability as an opportunity to improve upon existing systems and develop more secure solutions. In contrast, the late majority and laggards may perceive the security vulnerability as a reason to delay or reject the adoption of such technology in healthcare. An interesting observation is the variation in technological aspects in each sector, as well as the expression of sentiment towards them. The difference in sentiment between industries may be useful in highlighting the distinct adoption patterns prevalent in each sector. By analysing these patterns, valuable insights into the factors that influence the adoption of new technologies can be obtained and used to tailor strategies accordingly. Understanding the nuances of sector-specific adoption patterns enables stakeholders to make better-informed decisions, thereby facilitating the successful integration of emerging technologies across multiple domains.

\section{Evaluation}\label{sec:evaluation}

To quantitatively test our hypothesis, a ground truth analysis was performed to validate that the sentiment captured by the text mining approach are comparable to the results given by independent human annotators when asked to label whether such texts positively or negatively impact their outlook towards adopting an emerging technology. We measured impact using three metrics:

\begin{itemize}
    \item \textbf{Positive} - The text has a positive impact on the reader. Given this information, they are now more likely to accept, integrate, and/or use the technology in their business or personal life.

    \item \textbf{Negative} - The text has a negative impact on the reader. Given this information, they now feel against integrating and using the technology in their business or personal life.
    
    \item \textbf{Neutral} - The text has no impact on the reader and they feel indifferent about the technology.
\end{itemize}

To facilitate the annotation task, a bespoke web-based annotation platform accessible via a web browser was implemented. This eliminated any installation overhead and widened the reach of annotators. Annotators were presented with instructions explaining the task's requirements and then with the platform's interface consisting of two panes. The first pane contained a randomly selected tweet to be annotated, as well as the referenced technology discussed in the tweet. The second pane contained the annotation choices for the aforementioned metrics.  

The crowdsourcing of labelling natural language often uses a limited number of annotators with the expectation that they are perceived to be experts \cite{tarasov2010using}. However, annotation is a highly subjective task that varies with age, gender, experience, cultural location, and individual psychological differences \cite{passonneau2008relation}. For example, Snow et al. \cite{snow2008cheap} investigate collecting annotations from a broad base of non-expert annotators over the Web. They show high agreement between the annotations provided by non-experts found on social media and those provided by experts. In this case, in this study, a crowdsourcing approach was adopted to annotate a randomly sampled set of 150 tweets (50 samples of positive, negative, and neutral tweets) by developing and disseminating a annotation platform on Twitter, enabling users to participate in the annotation process and contribute to the assessment of sentiment in the dataset. A total of 750 annotations were collected with five annotations per sample. A total of 20 independent annotators participated in the study.

To quantitatively measure the reliability of the collected annotations, we measured inter-annotator agreement using Krippendorff's alpha coefficient \cite{krippendorff2018content}. As a generalisation of known reliability indices, it was used as it applies to: (1) any number of annotators, not just two, (2) any number of categories, and (3) corrects for chance expected agreement. Krippendorff's alpha coefficient of 1 indicates perfect agreement, 0 indicates no agreement beyond chance, and -1 indicates disagreement. The values for Krippendorff’s alpha coefficient were obtained using Python's computation of Krippendorff's alpha measure \cite{kpython}. 

Krippendorff suggests $\alpha$ = 0.667 as the lowest acceptable value when considering the reliability of a dataset \cite{krippendorff2004reliability}. The inter-annotator agreement of the annotated dataset in this study was calculated as $\alpha$ = 0.769, with a total of 89 samples out of 150 (59.3\%) achieving full agreement. The relatively high agreement ($\alpha$ = 0.769) illustrates the relative reliability of the annotations which delineate the impact of the presented texts on technology users. 

To evaluate our proposed text mining approach against a human annotator perspective, annotated tweets were used to create a gold standard. For each tweet in the sample, an annotation agreed by the relative majority of at least 50\% was assumed to be the ground truth. For example, the tweet \textit{'Cyber attacks on the rise how secure is your router network'} was annotated with negative four times and once with neutral, thus negative was accepted as the ground truth. When no majority annotation could be identified, a new independent annotator resolved the disagreement. For example, the tweet \textit{'We may have soon pills or grain size sensors in US reporting in real time'} was annotated twice with positive, twice with neutral, and once with negative. Thus, the independent annotator accepted neutral as the ground truth. A total of three samples suffered from disagreement and were resolved by the independent annotator.

\begin{table}
\centering
\caption{Confusion matrix comparing text mining outputs with human annotations}
\label{tab:confusion_matrix}
\begin{tabular}{lrccc}
                                          & \multicolumn{1}{l}{}                   & \multicolumn{3}{c}{\textit{Human Annotations}}                                                                          \\ \cline{3-5} 
                                          & \multicolumn{1}{l|}{}                  & \multicolumn{1}{c|}{\textbf{Positive}} & \multicolumn{1}{c|}{\textbf{Negative}} & \multicolumn{1}{c|}{\textbf{Neutral}} \\ \cline{2-5} 
\multicolumn{1}{l|}{}                     & \multicolumn{1}{r|}{\textbf{Positive}} & \multicolumn{1}{c|}{37}                & \multicolumn{1}{c|}{0}                 & \multicolumn{1}{c|}{13}               \\ \cline{2-5} 
\multicolumn{1}{c|}{\textit{Text Mining}} & \multicolumn{1}{r|}{\textbf{Negative}} & \multicolumn{1}{c|}{2}                 & \multicolumn{1}{c|}{45}                & \multicolumn{1}{c|}{3}                \\ \cline{2-5} 
\multicolumn{1}{l|}{}                     & \multicolumn{1}{r|}{\textbf{Neutral}}  & \multicolumn{1}{c|}{8}                 & \multicolumn{1}{c|}{1}                 & \multicolumn{1}{c|}{41}               \\ \cline{2-5} 
\end{tabular}
\end{table}

The confusion matrix given in Table \ref{tab:confusion_matrix} shows how the sentiment categories are re-distributed when comparing the sampled dataset generated by the text mining approach with the gold standard formed using the collected annotations. Overall, 123 out of 150 samples (82\%) were in agreement. When considering the positive sentiment, 37 of the samples in the gold standard were in agreement with the text mining approach. Some instances were in disagreement, where 13 samples were categorised as neutral. No positive instance was in disagreement with the negative category. Of the 50 samples of negative tweets, 45 samples of the gold standard agreed, with 2 and 3 samples being in disagreement and annotated as positive and neutral respectively. Likewise, for neutral tweets, 41 samples were in agreement, with 8 and 1 instances being annotated as positive and negative respectively. Such disagreements illustrate the natural subjective nature of the task. Overall, the relatively high agreement between the impact of such texts on human annotators and the results generated by the text mining approach implies that the proposed automated method generates reliable results towards understanding the barriers and opportunities faced by technology adopters from large online corpora.

\section{Conclusion}\label{sec:conclusion}

This paper presents a scalable and automated framework towards tracking the likely adoption of emerging technologies. Such framework is powered by the automatic collection and analysis of social media discourse containing references to emerging technologies from a large landscape of adopters. In particular, to support the experiments presented herein, and subsequently remove the dependence on manual qualitative data collection and analysis, an automated text mining approach was adopted to compile a large corpus of over four million tweets covering five year's worth of data. Once pre-processed, the corpus was divided into datasets based on their publication month and year. To extract references to emerging technologies from text, a simple string-matching approach was applied to automatically identify tweets containing references to technologies that could be mapped to CyBOK's cybersecurity index. Under the hypothesis that the expression of positive sentiment infers an increase in the likelihood of impacting a technology user's acceptance to adopt, integrate, and/or use the technology, and negative sentiment infers an increase in the likelihood of impacting the rejection of emerging technologies by adopters, sentiment analysis was applied to extract the sentiment expressed towards the identified technology. For each technology, the sentiment polarity with the highest average score was used to determine the overall sentiment expressed during the specific month and year. 

Notably, this study reports that instances of negative sentiment disclose the obstacles faced by technologies as a result of the dissemination of false information or their participation in malicious activities. In the context of risk assessment, a crucial aspect of a company's decision-making process, this information can serve as an additional factor in assessing the risks, costs, and benefits an organisation may face upon deploying such technologies, including the overall security of their technology systems and data. In the context of the stages of technology adoption, these obstacles may contribute to delays in progressing through the adoption stages or even lead to the rejection of the technology altogether. On the other hand, the expression of positive sentiment is useful for recognising the benefits and advantages of adopting particular technologies, as it provides insight into how other organisations with similar structures have successfully integrated them. By refining the data, the framework can also be customised to concentrate on specific industries, such as education and healthcare, which enhances its flexibility and makes it a dynamic solution that can adapt to the unique needs of different sectors.

To quantitatively test our hypothesis, a ground truth analysis was performed to validate that the sentiment captured by the text mining approach are comparable to the results given by human annotators when asked to label whether such texts positively or negatively impact their outlook towards adopting an emerging technology. The collected annotations demonstrated comparable results to those of the text mining approach, illustrating that automatically extracted sentiment expressed towards technologies are useful features in understanding the landscape faced by technology adopters across various stages of adoption.

\section{Future Work}\label{sec:future work}

Given the positive results of this preliminary study, the next step is to investigate the use of context-aware sentiment extractors to reduce false positives in emerging technology sentiment analysis. While the current method has presented useful insights, it may occasionally misinterpret sentiments due to a lack of context awareness. For example, whereas one would classify the tweet \textit{`cyber attack quick response guide'} as expressing a neutral sentiment, VADER's results report that the tweet expresses a negative one due to the presence of the word \textit{`attack'}. By employing advanced techniques, such as sentiment analysis models based on deep learning, the accuracy and reliability of the findings presented here can be enhanced, allowing for a more nuanced comprehension of the factors influencing technology adoption.

In addition, the results presented herein may be used to support event prediction. In particular, narratives which surround the cybersecurity threats and attacks faced by emerging technologies, particularly dialogues that express a negative sentiment or other linguistic elements that describe the intent to disrupt or cause harm, may be monitored in real-time and subsequently aid the identification and prediction of activities such as the launch of a malicious cyber attack aimed towards a particular technology.

\bibliographystyle{unsrt}  
\bibliography{references}  






\end{document}